      [1999/12/01 v1.4c Il Nuovo Cimento]
\documentclass{cimento}

\usepackage{graphicx}  
\usepackage{dcolumn}
\usepackage{bm}
\def\Rapt{{R_{AA}(p_T)}}
\def\Ra{{R_{AA}}}
\def\v2{{v_{2}(p_T)}}

\title{Impact of temperature dependence of the energy loss\\ on jet quenching observables }
\author{F.~Scardina\from{ins:x},\from{ins:f}\ETC,
M.~Di Toro\from{ins:e},\from{ins:f},
V.~Greco\from{ins:e},\from{ins:f}}
\instlist{\inst{ins:x} Department of Physics, University of Messina, I-98166 - Messina
  \inst{ins:e} Department of Physics and Astronomy, University of Catania, Via S. Sofia 64, I-95125 Catania.
\inst{ins:f}INFN- Laboratori Nazionali del Sud, Via S. Sofia 62, I-95125 Catania
}

\PACSes{\PACSit{12.38.Mh, 25.75 -q , 25.75 Ld}
{}}
\begin{document}

\maketitle

\begin{abstract}
The quenching of jets (particles with $p_T>>T, \Lambda_{QCD}$) in ultra-relativistic heavy-ion collisions
has been one of the main prediction 
and discovery at RHIC. We have studied, by a simple jet quenching modeling, the correlation between different observables like the nuclear modification
factor  $\Rapt$, the elliptic flow $v_2$ and the ratio of quark to gluon suppression $R_{AA}(quark)/R_{AA}(gluon)$. We show that the
relation among these observables is strongly affected by the temperature dependence of the energy loss. In
particular the large $v_2$ and and the nearly equal $\Rapt$ of quarks and gluons can be accounted for only if
the energy loss occurs mainly around the temperature $T_c$ and the flavour conversion is significant.Finally we point out that the efficency in the conversion of the space eccentricity into the momentum one ($v_2$) results to be quite 
smaller respect to the one coming from elastic scatterings in a fluid with a viscosity to entropy density
ratio $4\pi\eta/s=1$.
\end{abstract}

\section{Introduction}
The experiments at the Relativistic Heavy Ion Collider (RHIC) have given clear indications of  the formation
of Quark Gluon Plasma (QGP). One way to probe the QGP is to exploit the high energy jets ($p_T >> T,
\Lambda_{QCD}$) produced by the hard collisions at the initial stage. They are internal probes
propagating through the
fireball and interacting  with the medium losing energy, hence carrying information on its properties as
proposed long ago in Ref.s \cite{Gyulassy:1990ye}. 
This energy loss can be quantified by the suppression of observed hadron
spectra at high transverse momenta $p_T$, namely $\Rapt$
\cite{rhic_white_paper}.
Altough the observation of the jet suppression cannot be questionated
there are several fundamental questions that still remain open.To investigate them we have constructed a model
to study two observables beyond the
$\Rapt$. One is the elliptic flow $v_2(p_T)$, that gives a measure of the angular dependence of
quenching, and the other is  $\Ra(q)/\Ra(g)$ that determines the flavor dependence of the suppression.
We suggest that the study of the correlation between $\v2$ and  $R_{AA}(q)/R_{AA}(g)$ carry information
on the temperature dependence of the quenching and 
on the mechanism of parton flavor conversion.We find that an energy loss that increase as $T\rightarrow T_c$,
the $q \leftrightarrow g$ in-medium conversion and an expansion-cooling of the fireball according to a
lattice QCD EoS  improve the agreement with the experimental data.
Even if the T dependence that has to be considered in the present modeling including only path-lenth
energy loss appear to be too extreme.

\section{Modeling the jet quenching}
\begin{figure}	
\includegraphics[width=6.5 cm, height=5 cm]{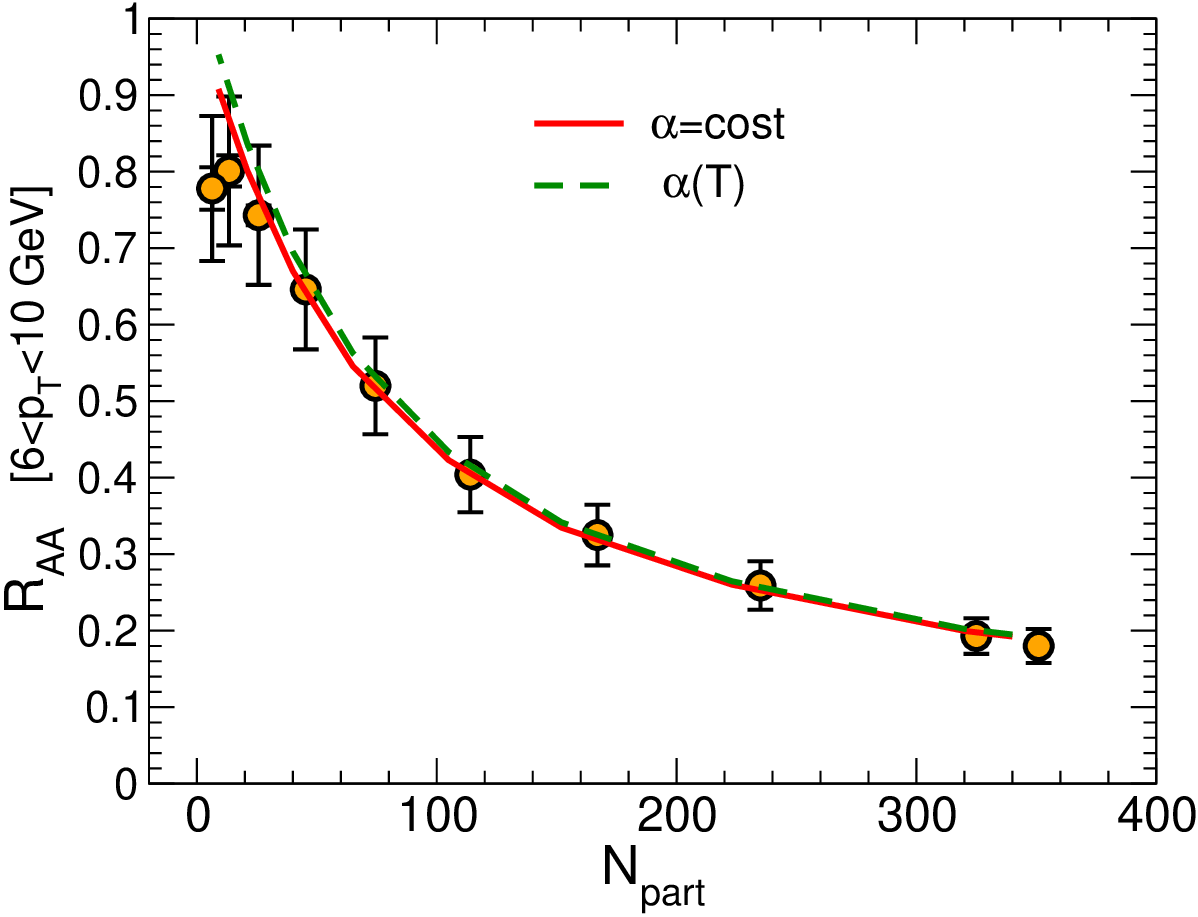}
\hspace {0.2 cm}
\includegraphics[width=6.3 cm, height=4.9 cm]{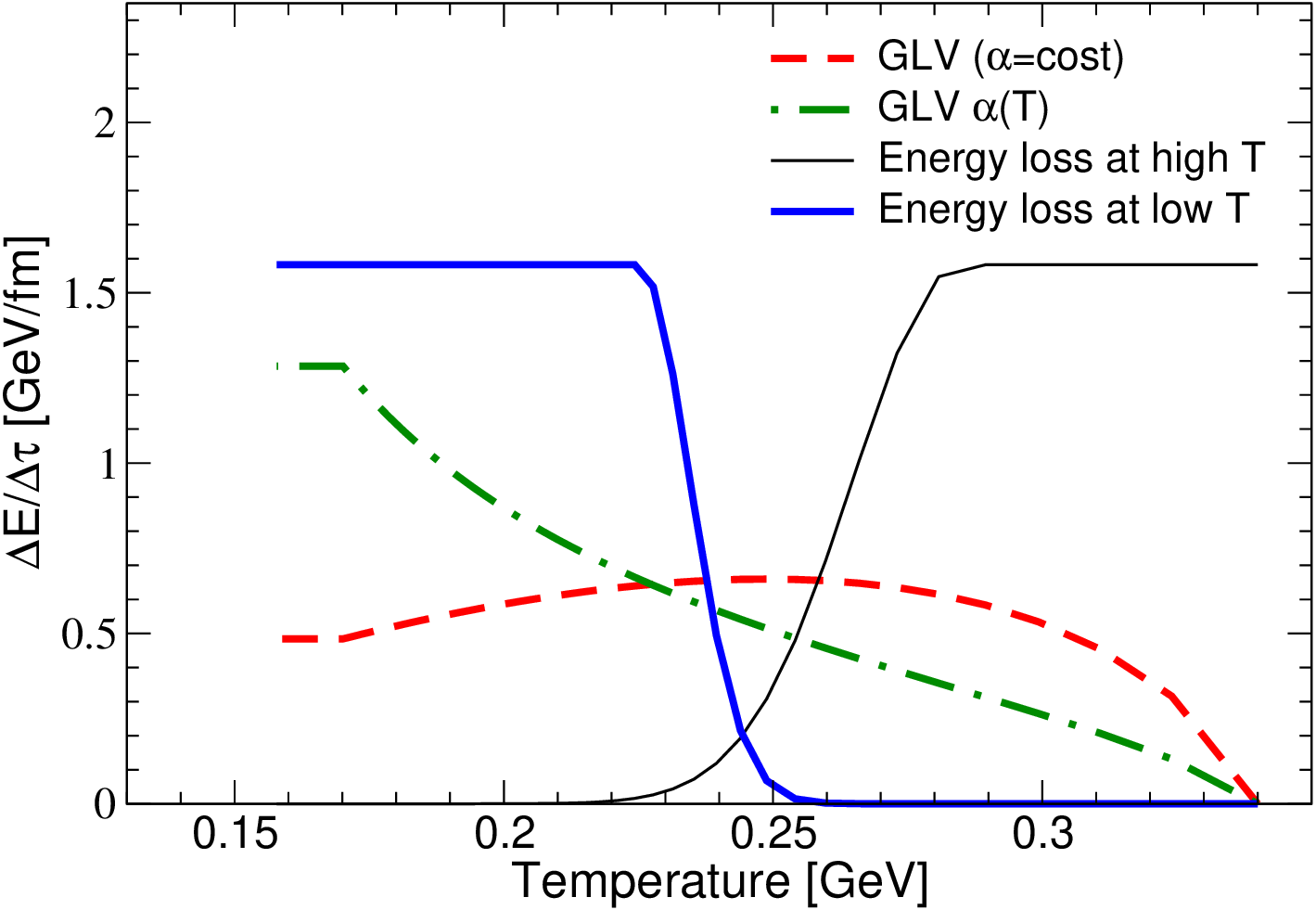}
\caption{Left panel: Nuclear Modification Factor as a function of the number of participant 
in $Au+Au$ at 200 AGeV \cite{Phenix_more}. Right panel: Temperature dependence of the energy loss for a parton with transverse momentum $p_T$ equal to 10 GeV. 
Dashed and dot-dashed lines represent the GLV energy loss with constant and T dependent $\alpha_s$ coupling (see
text). The thick line is the case in which the energy loss
takes place only closer to the phase transition and the thin line represents an opposite case
in which the energy loss take place only at high temperature T.}
\label{fig-RAA-centr}
\end{figure}

In the model the density profile of the bulk is given by the standard Glauber model while the hard parton
distributions in momenta space are calculated in the next-to-leading- order (NLO) pQCD scheme. For the
hadronization the Albino-Kramer-Kniehl (AKK) fragmentation functions  have been employed.
For further details see Ref. \cite{Sensitivity:2010}. With regard to the jets energy loss we have
employed various schemes, however to make a connection to the large amount of effort to
evaluate gluon radiation in a pQCD frame, we have used also the Gyulassy-Levai-Vitev (GLV) formula at first order in the opacity
expansion \cite{Gyulassy:2000gk,Gyulassy:2003mc}:

\begin{eqnarray}
\frac{\Delta E(\rho,\tau,\mu)}{\Delta \tau}=\frac{9\pi}{4}C_R \alpha_s^3 \rho(x,y,\tau)\,
\tau log\left(\frac{2E}{\mu^2\tau}\right)
\label{eq.glv}
\end{eqnarray}

where $C_R$ is the Casimir factor equal to 4/3 for quarks and 3 for gluons, $\alpha_s$ is the strong coupling,  $\rho(x,y,z)$ is the local density, $\tau$ the proper time, $E$ is the energy of the jet and $\mu=g T$ is the 
screening mass.
There are corrections to Eq.(\ref{eq.glv}) coming from higher order
that can be approximately accounted for by a rescaling Z factor of the energy loss. However this is not really
relevant for the
objectives of the present work because we will renormalize the energy loss in order to have the
measured amount of suppression $\Rapt$ for central collisions.
Usually in the GLV, as well as in other approaches, the temperature evolution of the strong
coupling $\alpha_s$ is discarded. We will consider the impact of such a dependence
to understand the amount of $T$ dependence coming simply from the asymptotic freedom.
In the right panel of Fig.\ref{fig-RAA-centr}  we show by dot-dashed and dashed lines the temperature dependence of the
energy loss
for the GLV with a dependence of the coupling(GLV-$\alpha_s(T)$) and with
a constant coupling $\alpha_s=0.27$ (GLVc). Furthermore we  show two other opposite cases for $\Delta E/\Delta \tau$:
the thick line
that shifts the energy loss to lower temperature (hence low density $\rho$ or entropy density $s$)
as suggested in \cite{Pantuev:2005jt,LS_PRL} and the other (thin line) that gives a dominance of
quenching at high $T$, considered here just for comparison respect to the opposite case.
We have applied our modelling of the jet quenching to Au+Au collisions at 200 AGeV
and in the left panel of Fig.\ref{fig-RAA-centr}  we can see that, once the $\Rapt$ at $0-5\%$ is fixed, the dependence  on centrality is correctly predicted with a 
GLV formula for both constant and T-dependent $\alpha_s$.
Hence looking at $\Ra$ one is not able to clearly discriminate the
temperature dependence of the quenching and not even the details of the density profile
\cite{Sensitivity:2010}.

\section{Angular and Flavor dependence of the Quenching}  
\begin{figure}
\includegraphics[width=6.5cm,height=5.0cm]{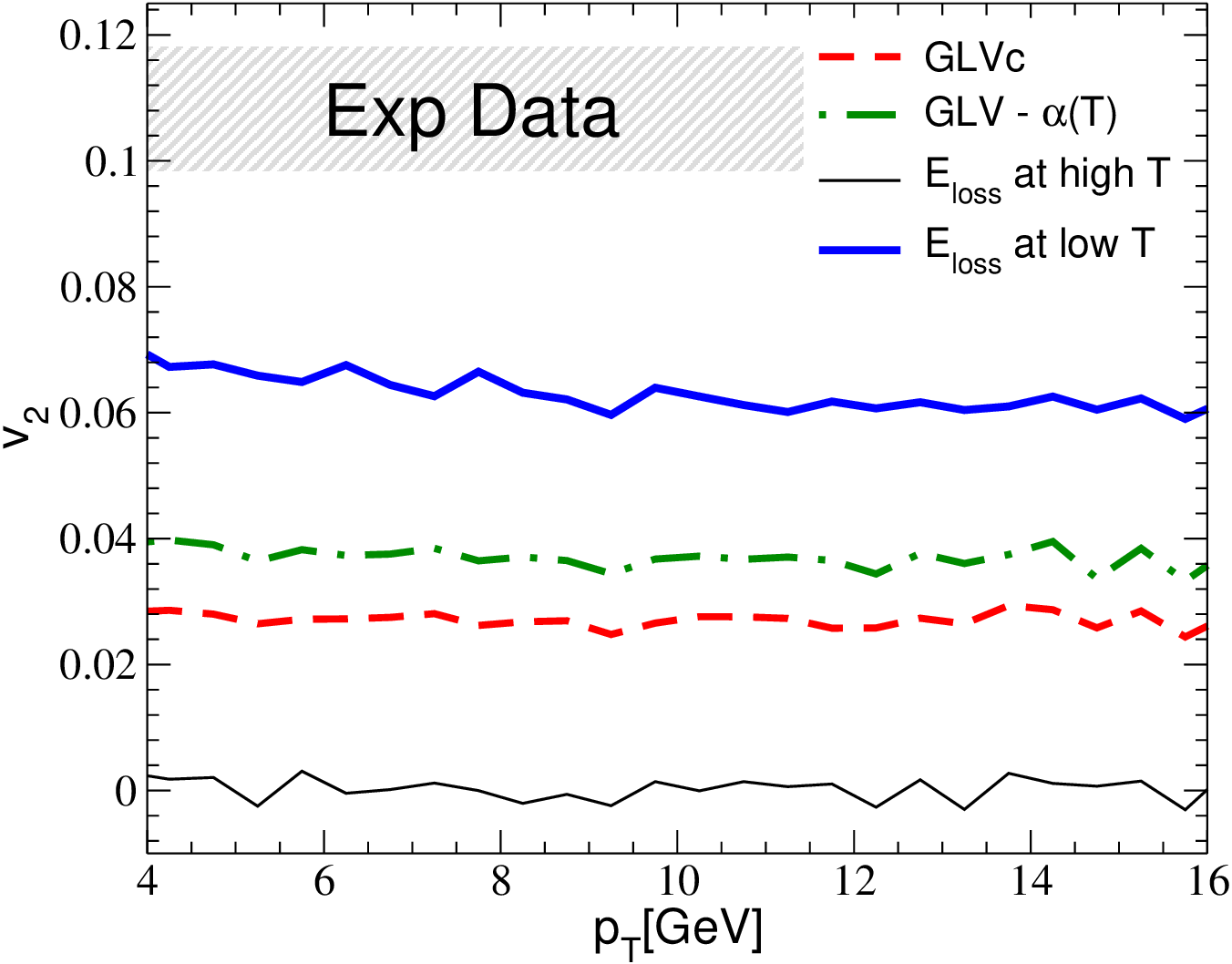}
\hspace {0.2 cm}
\includegraphics[width=6.5cm,height=5.0 cm]{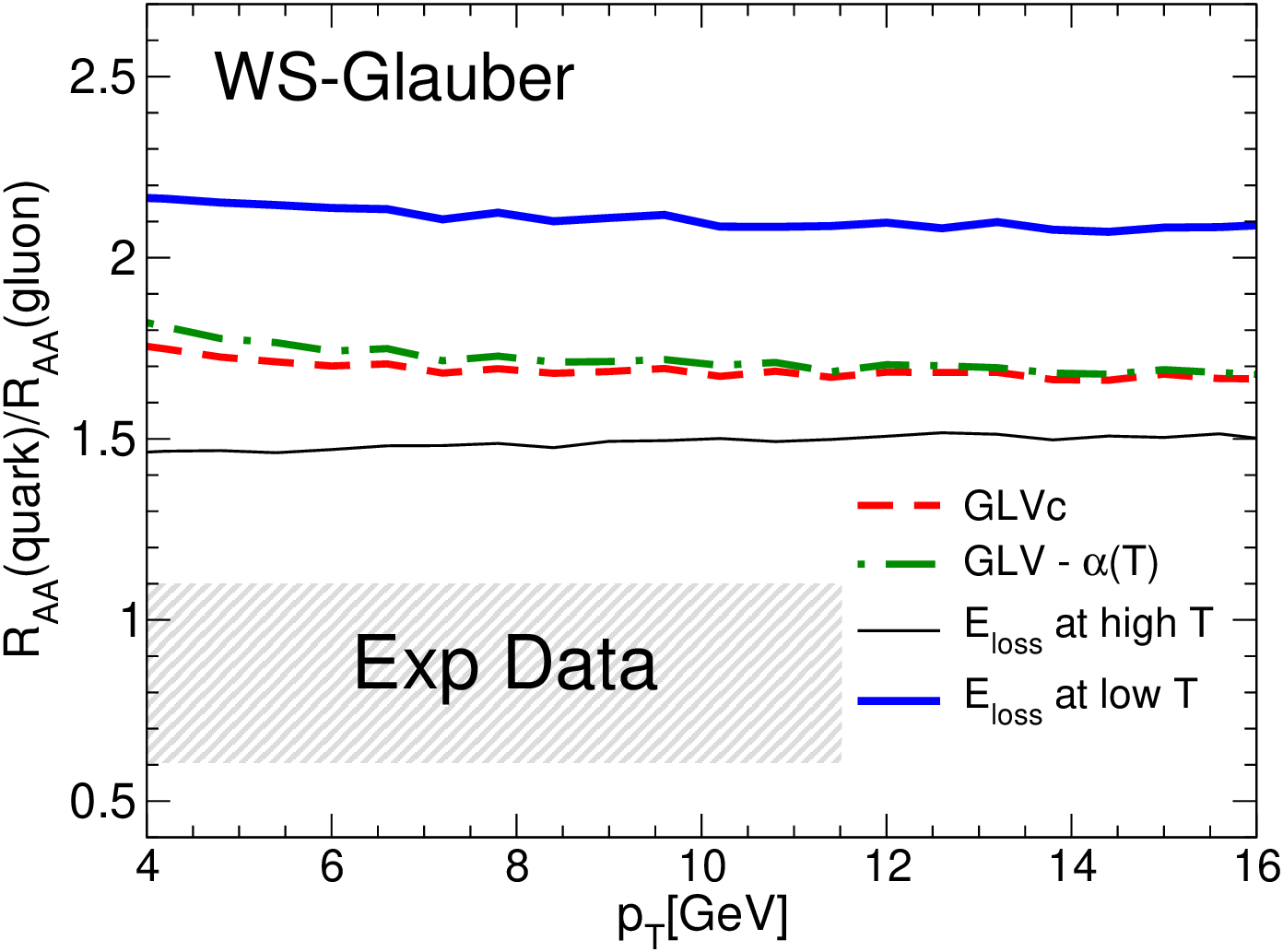}
\caption{Left panel: Elliptic Flow for pions (b=7.5 fm) coming from quark and gluon fragmentation for the
different T dependence
of the energy loss as shown in the right panel of Fig.\ref{fig-RAA-centr}.The shaded area shows the experimental data\cite{Phenix_more,Phenix_more2}. Right panel: Ratio of quark to gluon $R_{AA}$ for the
different T dependence of the energy loss as shown in the right panel of
Fig.\ref{fig-RAA-centr}. The shaded area
approximatively shows the value expected for the ratio according to experimental observations and using the
AKK fragmentation function } 
\label{v2_wood}
\vspace{0.5cm}
\end{figure}

Generally, jet quenching modeling has not been able to simultaneously describe $R_{AA}$ and the elliptic flow
$v_2$. In particular, experimental data relative to $v_2$ are
 considerably larger than theoretical prediction. We have explored the relation between the temperature
dependence of quenching and the value of the elliptic flow and 
in the left panel of Fig.\ref{v2_wood} we can see that even if the amount of total quenching has been fixed to the experimental
value of $\Rapt$ the amount of elliptic flow is strongly dependent on the 
temperature dependence of the $E_{loss}$. 

This correlation is due to the fact that at variance
with $\Rapt$ the $\v2$ has a longer formation time because the jets have to
explore the shape of the fireball to realize its asymmetry.Therefore it seems to be quite likely that experiments are telling us that quenching does not take place mainly at the very early time (temperature) of the collision but mainly at later times close to the phase transition.This would mean that the quenching is not proportional to the density (or entropy density) but a
decreasing function of it with a maximum at $T\sim T_c$ . This is what is essentially discussed also 
in Ref.s\cite{Pantuev:2005jt,LS_PRL} 
where however it was implicitely assumed that the amount of quenching
of quarks and gluons are equal among them and to the hadronic one. Here we have modified such assumptions
showing that the temperature (or entropy density) dependence of $E_{loss}$ modifies not only the
$\v2$ but also the relative amount of quenching of quarks and gluons.

\subsection{Quark to Gluon Modification Factor}
\begin{figure}	
\includegraphics[width=6.5cm, height=5cm]{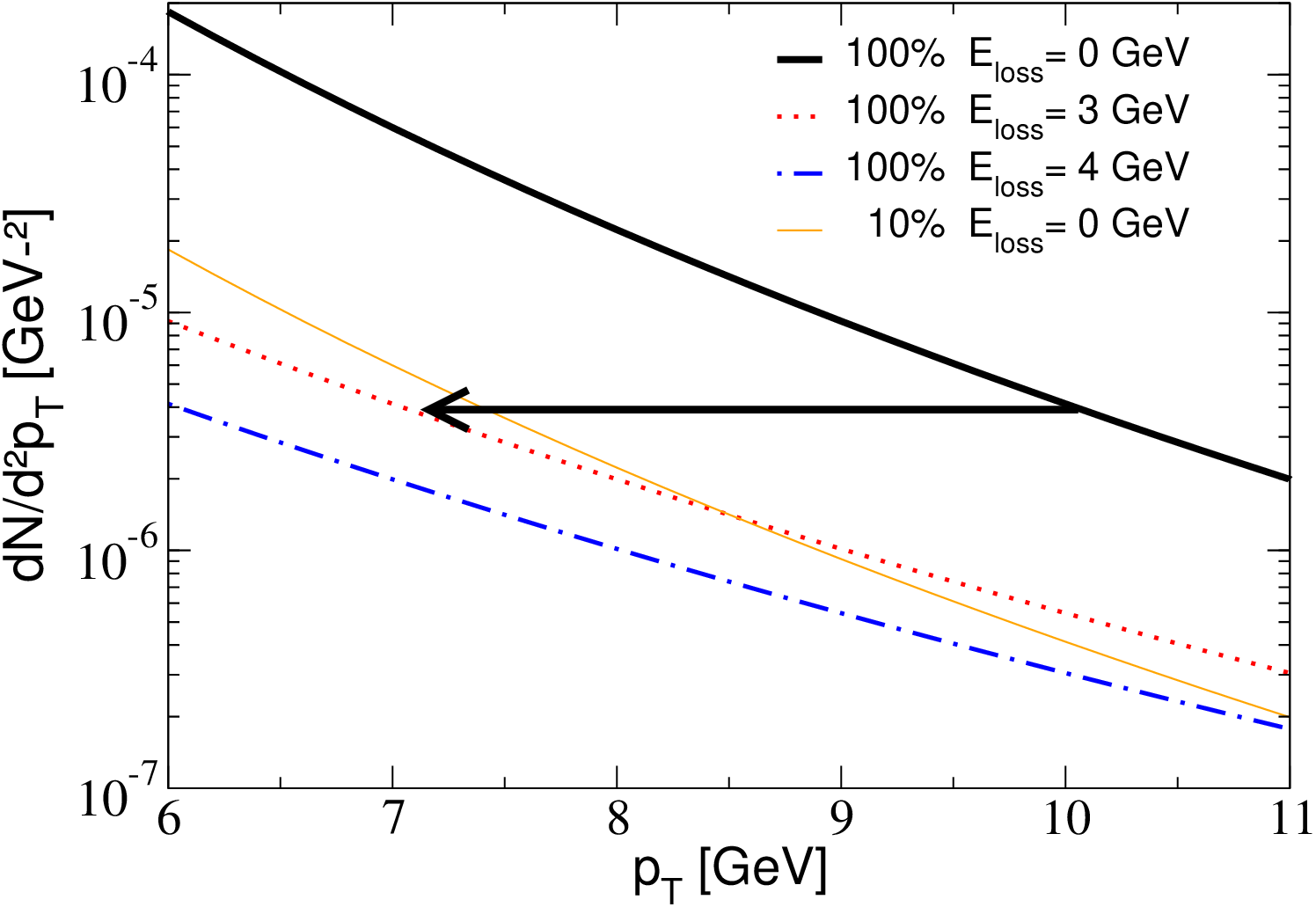}
\hspace {0.2 cm}
\includegraphics[width=6.5cm,height=5cm]{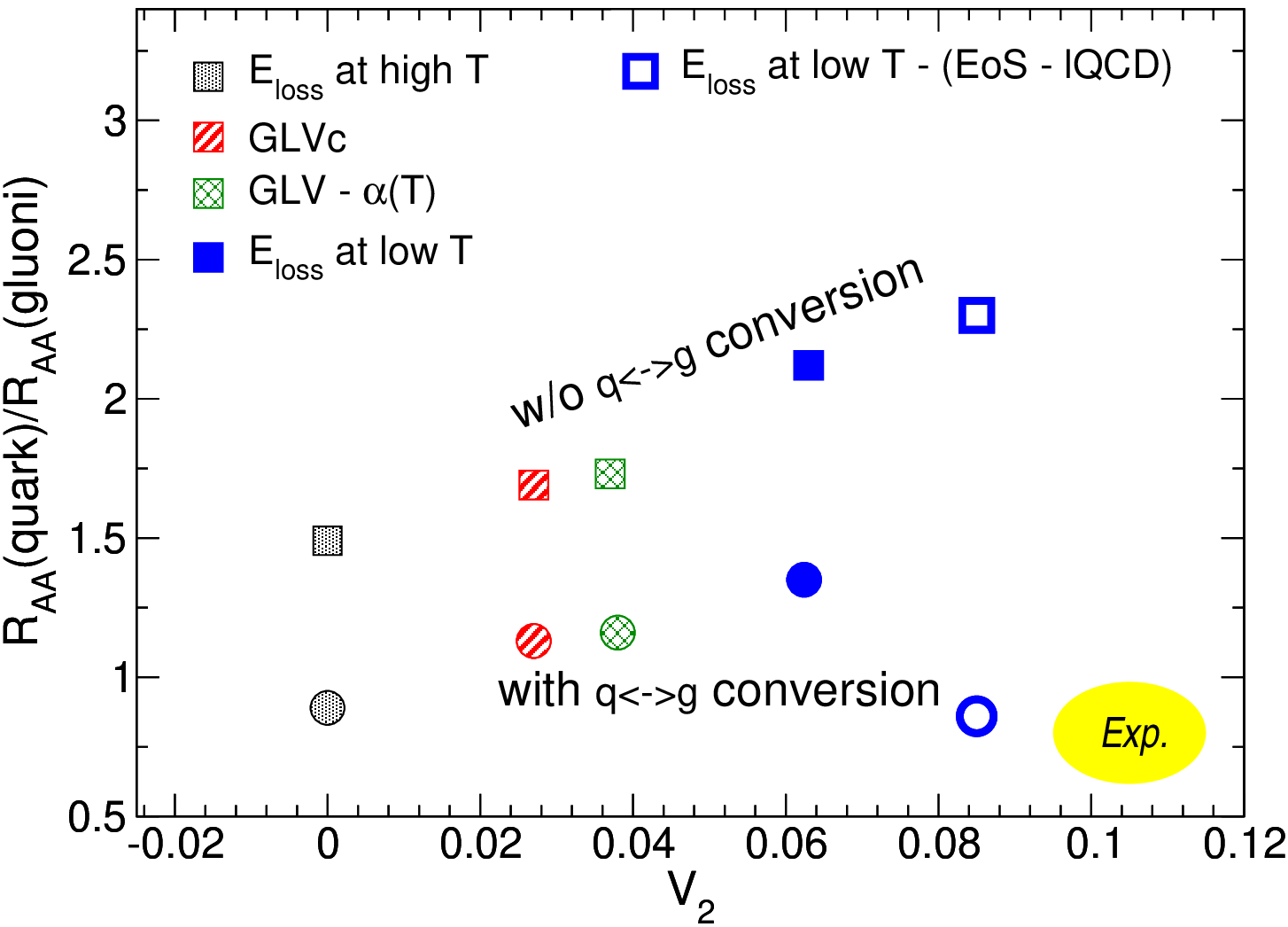}
\caption{Left panel: Spectra for partons that lose a fixed amount of energy (see text). Right panel: Correlation between the $R_{AA}(q)/R_{AA}(g)$ and the elliptic flow of pions with transverse momentum in the range $6<p_T (GeV) <10$.
The squares refer to calculation without jet conversion while the circle  are the values obtained including  jet
conversion with a $K_c=6$ factor. }
\label{fig-Rapporto-deltaE}
\end{figure}

Due to its SU(3) Lie algebra the energy loss of gluons is $9/4$ larger than the quark one. For this reason sometimes it is assumed that the ratio between
the quark and gluon suppression ($\Ra(q)/\Ra(g)$) is equal to $9/4$.
From this  one would think that the (anti-)protons are more suppressed respect
to pions because they
come more from gluon fragmentation than from quarks fragmentation respect to pions, at least according to the
AKK fragmentation function we employ. 
The data at RHIC however have shown that even outside the region where coalescence should be
dominant \cite{Fries:2008hs,Greco:2007nu} the protons and the antiprotons appear to be less suppressed than
the pions and $\rho^0$ \cite{Sickles_QM09,raa-flavor}. 
Again we can see that going beyond the simple amount of quenching given by $\Rapt$ both the
azimuthal dependence and the flavor dependence of the quenching appear
to be in disagreement with the data. 
We call this open issues the "azimuthal" and the "flavor" puzzle respectively.
We will show that even if $\Ra$ for central collisions is fixed to be $\sim 0.2$ the
$\Ra(q)/\Ra(g)$ is significantly affected by the temperature dependence of $E_{loss}$.\\
In the right panel of Fig.\ref{v2_wood} we show the ratio of the $\Ra(q)/\Ra(g)$ for  four
different temperature dependences of the energy loss $E_{loss}$, as in Fig.\ref{fig-RAA-centr} (right).
We can see that the standard GLVc energy loss does not give the expected ratio $9/4$ for $\Ra(q)/\Ra(g)$
but a lower value, around 1.8, which represents already a non negligible deviation from 2.25.
We can however see that if the energy loss would be strongly T dependent and dominant in the $T\sim T_c$ region $\Ra(q)/ \Ra(g)$ can increase up to about 2.2 on the contrary, if it is dominant in the high temperature region (thin solid line)
the $\Ra(q)/\Ra(g)$ can become as small as 1.5.\\
To understand this behavior is useful to consider the left panel of  Fig.\ref{fig-Rapporto-deltaE} where the transverse momentum distribution of initial parton and those obtained if all partons lose the same amount of energy (3 GeV and 4 GeV in the dotted and dot-dashed line, respectively) are shown. The effect of the quenching
in this oversimplified case is to shift the spectra by a quantity equal to the amount of energy loss in a way
indicated by the black arrow. Because of the rapid falling distribution the spectra after quenchig are in
these cases one order of magnitude smaller respect to the initial one. Therefore the $10\%$ of the spectrum
without quenching, indicated by the thin line, is comparable to the spectrum in the case of $E_{loss}$= 3-4 GeV . This means that if there are particles  that lose a very small quantity of energy,
like in the case of quenching at high temperature, they strongly influenced the final spectra for both quarks
and gluons and damping the difference between the $R_{AA}$ of quarks and gluons. For energy
loss dominated by low temperature all particles lose energy and this increases the difference between the
respective $R_{AA}$. A similar effect could come not only from the T-dependence of $E_{loss}$ but from
a core-corona effect \cite{Becattini:2009}. For further explanations see Ref. \cite{Sensitivity:2010}.

\section{Correlation between $R_{AA}(q)/R_{AA}(g)$ and elliptic flow}
We have seen (Fig. \ref{v2_wood},left and right panel) that an energy loss predominant at low T move $v_2$ toward observed data but move also the ratio
$(\Ra(q)/\Ra(g)$ away from experimental indications. This is evident if one looks at the upper symbols of
Fig.\ref{fig-Rapporto-deltaE} (right)  where  $(\Ra(q)/\Ra(g)$  vs  $v_2)$ is shown. To solve the "flavor
puzzle" inelastic collisions 
that cause a change of the flavor has been invoked \cite{Liu:2006sf,Liu:2008zb,Liu:2008bw}. 
Such a process would at the end produce a net conversion of quarks into gluons. Hence a
decrease of gluon suppression respect to the original suppression and an increase of the quark one. 
In Ref.\cite{Liu:2006sf} it has been calculated the conversion rate of a 
quark jet to a gluon jet and vice versa due to two-body scatterings. An enhancement factor $K_c=4-6$
that accounts for non-perturbative effect is needed to produce a nearly equal suppression of
quarks and gluons.We have included such a mechanism in our model. The results are the lower symbols in the right panel of Fig 3.
We can see that the $q\leftrightarrow g$ conversion does not affect the $v_2$ and allows to get closer to the
experimental observed value,i.e. a $v_2 \sim 0.1$
and an $\Ra(q)/\Ra(g) \leq 1$ (to account for the $\Ra(p+\bar p)>\Ra(\pi^++\pi^-)$ with AKK fragmentation
function).


\section{Impact of the Equation of State}
As a last point, we have observed that if the quenching is predominant near the phase transition the question
of the corret equation of state arises. In fact the free gas approximation is no longer a reasonable
approximation just close to $T_c$.
We have made some explorative studies on the impact that a more correct  equation of
state (EoS)can have on the correlation between $\Ra(q)/\Ra(g)$ and 
$v_2(p_T)$. 
Making a fit to the lattice QCD data \cite{lQCD-EoS} we have obtained the following relation  between density and temperature 
\begin{equation}
\frac{T}{T_0}=\left(\frac{\rho}{\rho_0}\right)^{\beta(T)}
\label{rho-t-lQCD}
\end{equation}
where  $\beta(T)=1/3-a(T_c/T)^n$ with $T\ge T_c$, $a=0.15$ and $n=1.89$ and of course for $T>> T_c$ one gets $\beta \sim 1/3$.
In order to estimate the impact of this correction we have performed a simulation for the 
$\Delta E_{loss}(T)$ behavior represented by the thick solid line in the right panel of Fig.\ref{fig-RAA-centr},
which is similar to the delayed energy loss proposed by Pantuev \cite{Pantuev:2005jt} 
as a solution for the observed large elliptic flow. 
We consider only this case because it is of course the one that is much more affected
by the modification implied by Eq.(\ref{rho-t-lQCD}).
The results are given by open symbols in the right panel of Fig.\ref{fig-Rapporto-deltaE}.Respect to the free gas expansion cooling the system spends more time at $T\sim T_c$. This reflects in a further enhancement of both  $\v2$ and efficency of the $q \leftrightarrow g$ conversion moving the two observables closer to experimental data (shaded area in Fig.\ref{fig-Rapporto-deltaE} (right))  

\begin{figure}	
\includegraphics[width=6.5cm, height=5cm]{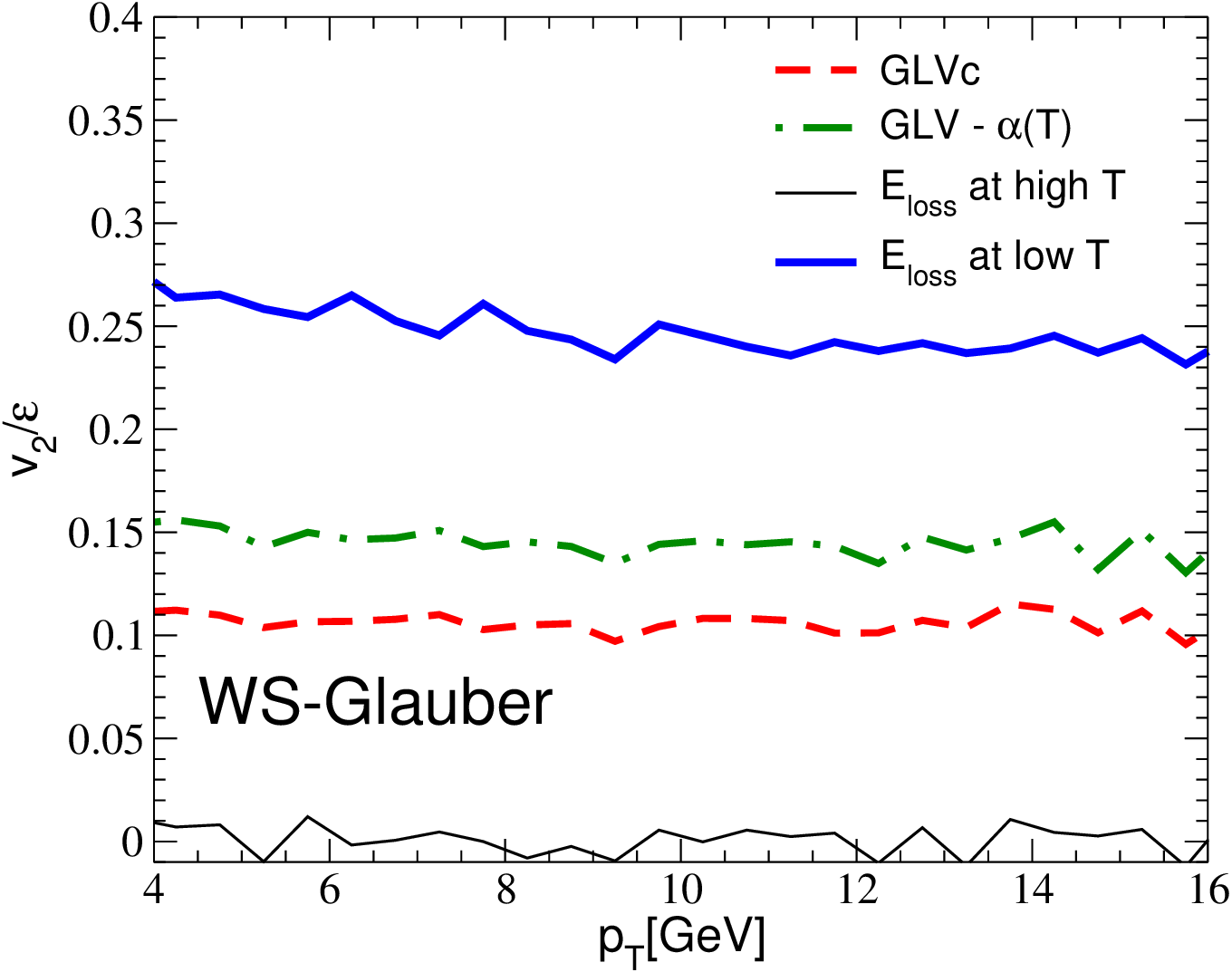}
\hspace {0.2 cm}
\includegraphics[width=6.5cm,height=5cm]{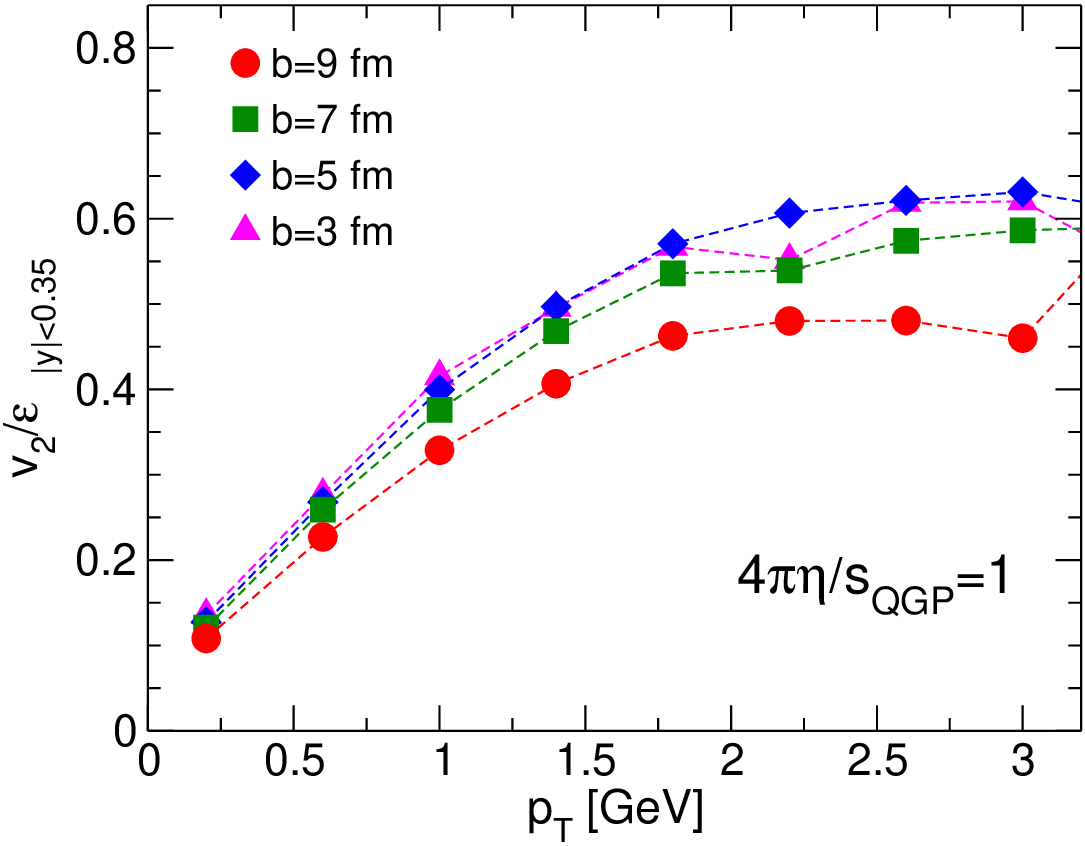}
\caption{Left panel: Ratio between the elliptic flow and the eccentricity  
 $v_2/\epsilon$ as a function of the transverse momentum $p_T$ of pions produced in collision with impact parameter equal to $7.5$ fm for different temperature dependence of the energy loss, as shown in the right panel of Fig. \ref{fig-RAA-centr}. Right panel: $v_2/\epsilon$ as a transverse momentum function estimated in a parton cascade approach Ref.\cite{Greco:2009ds} at different impact
parameters}
\label{v2_eccenticit}
\end{figure}

As a last point we want to mention the issue of the efficiency of conversion of initial spatial asimmetry
$\epsilon$
into a $v_2$, namely the $v_2/\epsilon$. In the left panel of Fig.\ref{v2_eccenticit}  we show the different values of $v_2/\epsilon$ that
one obtains through the simple path-length mechanism studied with our modeling, noticing that
it is at maximum $v_2/\epsilon \leq 0.25$. In the right panel of Fig. \ref{v2_eccenticit} we show the $v_2/\epsilon$ coming from elastic
scattering in a cascade approach for a fluid at finite shear viscosity to entropy density $4\pi\eta/s=1$
\cite{Greco:2009ds}.
Even if at slightly lower $p_T$ the $v_2/\epsilon$ in this case is about a factor of two larger. This seems
to indicate that a proper treatment of elastic energy loss can give an important constribution at least
at intermediate $p_T \sim 4-6$ GeV where presently the experimental data on $v_2$ are available.

\section{Conclusions}
We have pointed out the impact of peculiar temperature dependences of the energy loss 
on the elliptic flow  and on the ratio between the quark and gluon suppression  and their correlation. Moreover we have spot the relevance
that the $EoS$ may have in case of $E_{loss}$ dominant in the $T\sim T_c$ region.
In any case our study, although already revealing several interesting indications,
is mainly explorative and it can be considered as a benchmark. A more quantitative analysis should be performed
with more sophisticated models that include the energy loss fluctuations,
realistic gain and loss processes, elastic energy loss and a more accurate description of the bulk.
Furthermore it should be explored if mass dependent formation time can affected the correlation between $v_2$
and $\Ra(q)/\Ra(g)$ \cite{Markert}.
Obviously,it is also important to study how the longer lifetime and higher temperatures which will be reached
at LHC energies could affect the observed correlations.

\end{document}